# A QUANTUM-DRIVEN-TIME (QDT) QUANTIZATION OF THE TAUB COSMOLOGY

Arkady Kheyfets
Department of Mathematics
North Carolina State University
Raleigh, NC 27695-8205

Warner A. Miller
Theoretical Division
Los Alamos National Laboratory
Los Alamos, NM 87545

19 September 1994

## Abstract

We present here an application of a new quantization scheme. We quantize the Taub cosmology by quantizing only the anisotropy parameter $\beta$ and imposing the super-Hamiltonian constraint as an expectation-value equation to recover the relationship between the scale factor $\Omega$ and time $t$. This approach appears to avoid the problem of time.

# 1 Quantum-Driven Time (QDT) Quantization

We present here an application of a new approach to quantum gravity wherein the very concept of time emerges by imposing the principle of general covari-ance as weakly as possible.[1] In particular, the four constraints are imposed as expectation-value equations over the true dynamic degrees of freedom of the gravitational field – a representation of the underlying anisotropy of the 3-space. In this way the concept of time appears to be inextricably intertwined and woven to the initial conditions as well as to the quantum dynamics over the space of all conformal 3-geometries. This quantum-driven-time (QDT) approach will ordinarily lead to quantitatively different predictions than either the Dirac or ADM quantizations, and in addition, our approach appears to avoid the interpretational conundrums associated with the problem of time in quantum gravity.[2]

In this short paper we apply this QDT approach to the Taub cosmology [4] and numerically solve the QDT Schrödinger equation.

## 2  QDT Quantization of the Taub Cosmology

The Taub cosmology is an axisymmetric homogeneous cosmology parameterized by a scale factor $\Omega(t)$, and an anisotropy parameter $\beta(t)$. The line element may be expressed as

$$ds^2 = -dt^2 + a_o^2 e^{2\Omega} \left(e^{2\beta}\right)_{ij} \sigma^i \sigma^j, \qquad (1)$$

where $(\beta) = diag(\beta, \beta, -2\beta)$, and the one forms as, $\sigma^1 = \cos\psi d\theta + \sin\psi d\phi$, $\sigma^2 = \sin\psi d\theta - \cos\psi d\phi$, and $\sigma^3 = d\psi + \cos\theta d\phi$. The scalar 4-curvature is expressed in terms of $\Omega$ and $\beta$ and yields the action,

$$I_c = \frac{3\pi a_o^3}{4} \int \{(\dot\beta^2 - \dot\Omega^2) - \frac{1}{6}{}^{(3)}R\} e^{3\Omega} dt. \qquad (2)$$

Here ${}^{(3)}R = \frac{e^{-2\beta}}{2a_o^2 e^{2\Omega}}(4 - e^{-6\beta})$ represents the scalar 3-curvature.

We treat here the scale factor $\Omega(t)$ as the many-fingered time parameter [5] and the anisotropy $\beta(t)$ as the dynamic degree of freedom.[3] The momentum conjugate to $\beta$ is obtained from from the Lagrangian, $\mathcal{L}$.

$$p_\beta = \frac{\partial \mathcal{L}}{\partial \dot\beta} = \frac{3}{2}\pi a_o^3 e^{3\Omega} \dot\beta = m\dot\beta \qquad (3)$$

The QDT Hamiltonian for this cosmology can be expressed in terms of this momentum and the Lagrangian,

$$\mathcal{H}_{DYN} = p_\beta \dot{\beta} - \mathcal{L} = \frac{1}{2m}p_\beta^2 + \underbrace{\frac{m}{2}\left(\dot{\Omega}^2 - \frac{1}{6}{}^{(3)}R\right)}_{V}. \tag{4}$$

In the classical theory $\mathcal{H}_{DYN}$ can be used to construct either the Hamilton-Jacobi equation or the two Hamilton equations; however, to complete the dynamics (i.e. to restore general covariance) we must in addition impose the super-Hamiltonian constraint,

$$p_\beta^2 = m^2\left(\dot{\Omega}^2 + \frac{1}{6}{}^{(3)}R\right). \tag{5}$$

Using this QDT Hamiltonian (Eq. 4), the corresponding Hamilton-Jacobi equation and the standard quantization prescription we obtain the Schrödinger equation for the Taub cosmology.

$$-i\hbar\frac{\partial \Psi_s(\beta,t)}{\partial t} = -\frac{\hbar^2}{2m}\frac{\partial^2 \Psi_s(\beta,t)}{\partial \beta^2} + V\Psi_s(\beta,t) \tag{6}$$

The scale factor $\Omega$ in $V$ and $m$ in this equation should be treated as an unknown function of time, $t$. To complete the QDT quantization we impose, in addition, Eq. 5 as an expectation-value equation over $\beta$ (where $<\bullet>_s = \int \Psi_s^* \bullet \Psi_s d\beta$).

$$\dot{\Omega}^2 = \left(\frac{<p_\beta>_s}{m}\right)^2 + \frac{1}{6}<{}^{(3)}R>_s \tag{7}$$

This system of equations Eq. 6 and Eq. 7 provide us with a complete quantum-dynamic picture of the axisymmetric Taub model, and when augmented by the appropriate boundary conditions can be solved numerically.

We present a solution (Figs. 1 and 2) using an initially Gaussian wave packet.

$$\Psi_0 = Ce^{i(\beta-\beta_o)p_o/\hbar}e^{-(\beta-\beta_o)^2/\Delta\beta^2} \tag{8}$$

with $C^2 = \sqrt{2/\pi}/\Delta\beta$, $p_o = -100$, $\hbar = 1$, $\beta_o = 0$, $\Delta\beta = 0.1$ and $\Omega(t=0) = 1$. The solution was obtained using a 2nd-order split operator unitary integrator which preserves the norm exactly (up to round off errors).[6]

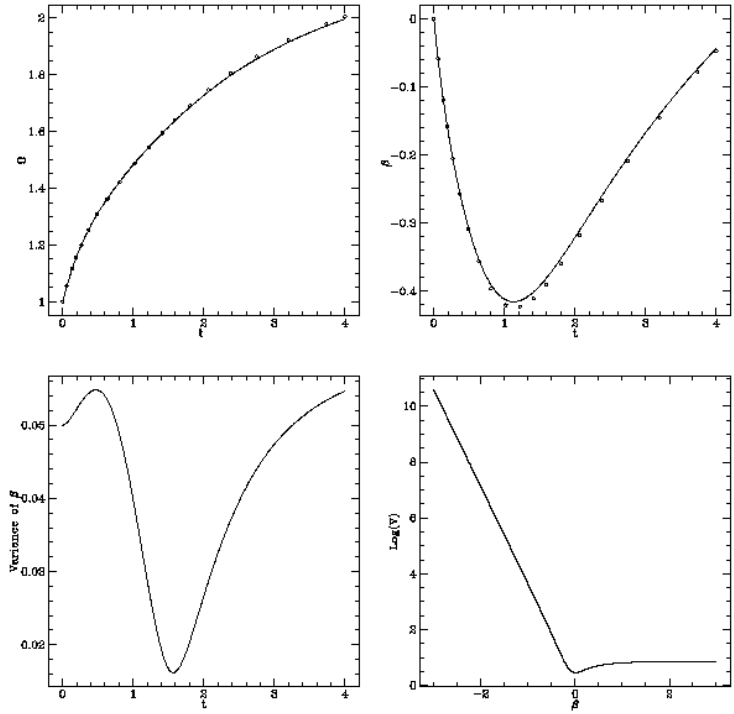

Figure 1: *The QDT quantization of the Taub cosmology.* The solid lines in each of the four graphs represent the quantum solution, while the small circular dots in the upper two graphs represent the classical trajectory. The scale factor's ($\Omega$) dependence on time ($t$) is shown in the upper left, while the expectation of the anisotropy ($<\beta>$) is displayed in the upper right. The lower left shows the variance in $\Psi_s$ throughout its evolution, and the lower right graph is a snapshot of the potential ($V$) at the end of the simulation.

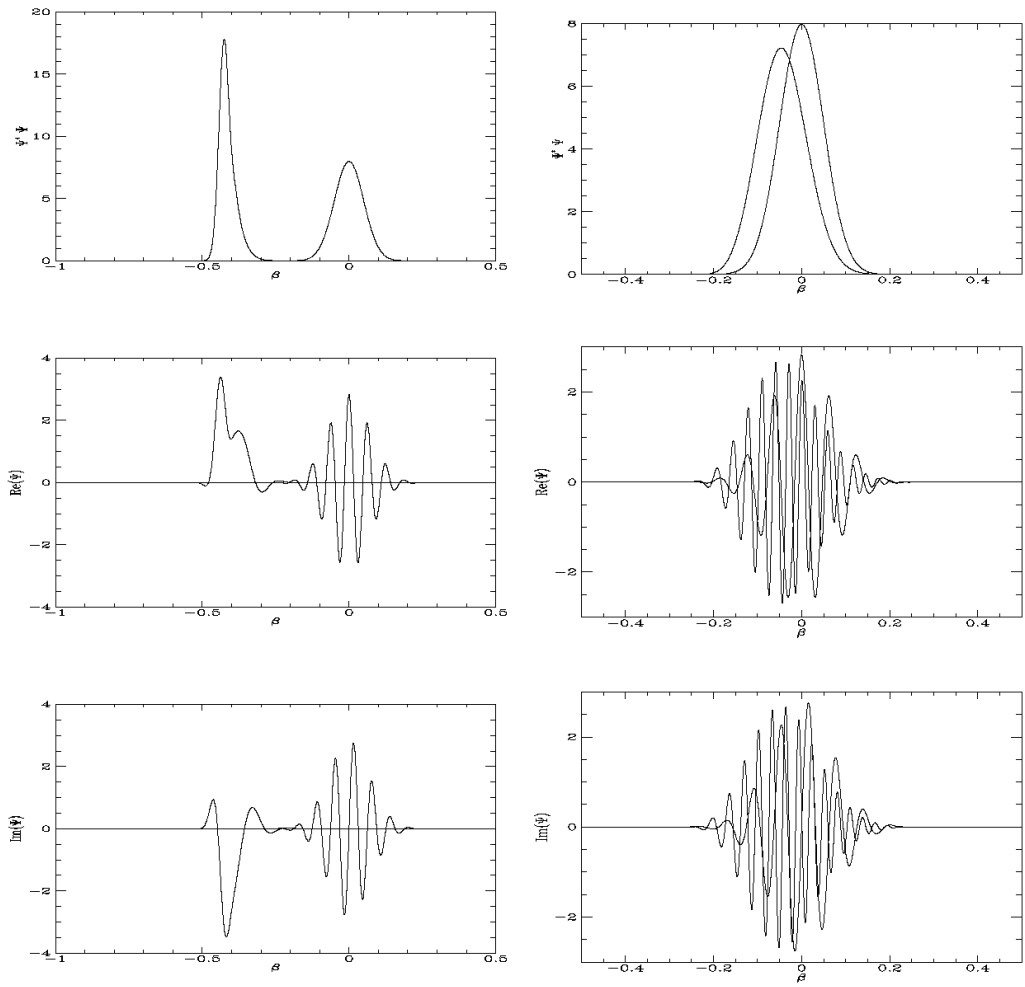

Figure 2: *The Taub wave function.* The left column of three graphs shows the initial ($t = 0$) wave function (right most) together with a snapshot of the wave function near the bounce ($t = 1\frac{5}{}$) (left-most wave function). The right column is similar but displays the right-going wave function well after the bounce at $t = 4$. The first column plots $\Psi_s^* \Psi$, while the second and third are the real and imaginary parts, respectively.

# 3 Time and Quantum Gravity

We have introduced in this paper an application of a new approach to the quantization of gravity (this approach need not be restricted to quantum cosmology). When the Taub cosmology was quantized our theory generated a quantum picture based on a post-ADM treatment[3] of the gravitational field dynamics and appears to be free from the conceptual difficulties related with time usually associated with both the Dirac and ADM quantization procedures. Only the dynamical part, the underlying anisotropy ($\beta$) has been quantized. The Hamiltonian participating in the quantization ($\mathcal{H}_{DYN}$) is not a square-root Hamiltonian. This absence of a square-root Hamiltonan is a generic feature of our QDT-quantization procedure.

We have demonstrated here using the Taub cosmology that the concept of time may very well be inextricably intertwined and woven to the initial conditions as well as to the quantum dynamics over the space of all conformal 3-geometries.

# Acknowledgements

We wish to thank J. Anglin, B. Bromley and R. Laflamme for their assistance, and especially S. Habib and P. Laguna for their help in solving the QDT equations numerically.